\documentclass [12pt]{article}
\usepackage{amsmath,amssymb,cite}

\numberwithin{equation}{section}

\begin{document}
\begin{flushright}
DIAS-STP-04-07
\end{flushright}
\begin{center}
{\bf Noncommutative $U(1)$ Gauge Theory As a Non-Linear Sigma Model}\\
\bigskip
{\bf Badis Ydri}
\bigskip

{\it School of Theoretical Physics, \\
Dublin Institute for Advanced Studies, Dublin, Ireland.}\\

\end{center}
\begin{abstract}
Noncommutative $U(1)$ gauge theory in $4-$dimensions is shown to be equivalent in some scaling limit to an ordinary non-linear sigma model in $2-$dimensions . The model in this regime is solvable and the corresponding exact beta function is found. We also show that classical $U(n)$ gauge theory on ${\bf R}^{d-2}{\times}{\bf R}^2_{\theta}$ can be approximated by a sequence of  ordinary $(d-2)-$dimensional Georgi-Glashow models with gauge groups $U(n(L+1))$ where $L+1$ is the matrix size of the  regularized noncommutative plane ${\bf R}^2_{\theta}$.

\end{abstract}
\section{Introduction}
The Moyal-Weyl noncommutative space is a $0-$dimensional matrix model and thus it is not a continuum manifold . It is  known that this space can be represented by infinite dimensional matrices  acting on some infinite dimensional Hilbert space\cite{nekrasov}. The fuzzy sphere on the other hand although it is a $0-$dimensional matrix model it acts on a finite dimensional Hilbert space \cite{madore,1,2,local}. In other words the fuzzy sphere can be represented by finite dimensional matrices .Clearly and in analogy with the continuum situation one should be able to get form one space to the other and vice versa. However and as it turns out we have  more structure in this case since the fuzzy sphere can in fact be thought of as a regularization of the noncommutative plane \cite{7}.

The noncommutative plane is given in terms of the algebra of the harmonic oscillator. The coordinates on this space are given by $x_1=\sqrt{\frac{\theta}{2}}(a+a^{+})$, $x_2=-i\sqrt{\frac{\theta}{2}}(a-a^{+})$ , $[x_1,x_2]=i{\theta}$. The algebra on the fuzzy sphere is given on the other hand in terms of the generators $L_a$ of the IRR $\frac{L}{2}$ of the $SU(2)$ Lie algebra. The global coordinates on the fuzzy sphere are defined by $
x_a=\frac{RL_a}{|L|}$ , $[x_a,x_b]=\frac{iR}{|L|}{\epsilon}_{abc}x_c$ , $\sum_ax_a^2=R^2$ . 

It is not difficult to argue from the above equations that the fuzzy sphere algebra is nothing else but a deformation of the Moyal-Weyl plane algebra which results in a finite dimensional Hilbert space\cite{mine}. Taking $L$ to infinity  reduces the fuzzy sphere to a noncommutative plane. This cut-off is gauge invariant as one can also see from the action. $U(1)$ gauge action on the noncommutative Moyal-Weyl plane is given by \cite{nekrasov}
\begin{eqnarray}
S_{\theta}=\frac{{\theta}^2}{4g^2}Tr\hat{F}_{ij}^2=\frac{{\theta}^2}{4g^2}Tr\sum_{i,j}\bigg(i[\hat{D}_i,\hat{D}_j]-\frac{1}{{\theta}^2}({\epsilon}^{-1})_{ij}\bigg)^2,\label{action1}
\end{eqnarray}
where $\hat{D}_i$ is the covariant derivative define by $\hat{D}_i=-\frac{1}{{\theta}^2}({\epsilon}^{-1})_{ij}\hat{x}_j+\hat{A}_i$ and $\hat{F}_{ij}$ is the curvature tensor. $Tr$ is over the infinite dimensional Hilbert space $H$ . This theory can be regularized by the following finite dimensional matrix model \cite{mine}
\begin{eqnarray}
S_{L,R}=\frac{R^2}{4g^2}\frac{1}{L+1}Tr_LF_{ab}^2=\frac{R^2}{4g^2}\frac{1}{L+1}Tr_L
\sum_{a,b}\bigg(i[D_a,D_b]+\sum_{c}\frac{1}{R}{\epsilon}_{abc}D_c\bigg)^2,\label{action2}
\end{eqnarray}
with the constraint \cite{nair,denjoeydri1,steinacker}
\begin{eqnarray}
D_aD_a=\frac{|L|^2}{R^2}~,~|L|^2=\frac{L}{2}(\frac{L}{2}+1).\label{constraint1}
\end{eqnarray}
The equations of motion are given by $2R[F_{cb},D_b]=i{\epsilon}_{abc}F_{ab}$ . They are solved by the zero-curvature condition $F_{ab}=0$ which are equivalent to $D_a=\frac{1}{R}L_a$ . This is exactly the fuzzy sphere \cite{jap1}. Expanding around this solution by writing $D_a=\frac{1}{R}L_a+A_a$ leads to $U(1)$ gauge theory on the fuzzy sphere . The above constraint is needed to describe a $2-$dimensional gauge field and also to stabilize the fuzzy sphere solution \cite{denjoeydri1}. This is also related to the fact that $a=1,2,3$ since the differential calculus on the fuzzy sphere is $3-$dimensional. $Tr_L$ is a finite dimensional trace over the Hilbert space $H_L$ , for example $Tr1=L+1$.

We are interested therefore in a continuum  double scaling limit of large $R$
and large $L$ taken together ( restricting the theory around the north pole for example ) as follows \cite{jap1,7}
\begin{eqnarray}
R,L~{\longrightarrow}{\infty}~;~{\rm
keeping}~\frac{R^2}{|L|^{2q}}=~{\rm
fixed}{\equiv}{\theta}^2~,~q={\rm real~ number}\label{limit}
\end{eqnarray}
The action (\ref{action2}) is seen to tend to
(\ref{action1}) with a resulting effective noncommutativity 
\begin{eqnarray}
{\theta}_{eff}^2={\theta}^2{\xi}^2~,~{\xi}^2=|L|^{2q-2}(L+1)\label{effe}
\end{eqnarray}
For $q>\frac{1}{2}$ ,
${\xi}^2{\longrightarrow}{\infty}$ when
$L{\longrightarrow}{\infty}$ and thus ${\theta}_{eff}$ corresponds
to strong noncommutativity. For $q<\frac{1}{2}$ we find that
${\xi}^2{\longrightarrow}0$ when $L{\longrightarrow}{\infty}$ and
${\theta}_{eff}$ corresponds to weak noncommutativity. For 
$q=\frac{1}{2}$ the effective noncommutativity parameter is
exactly given by ${\theta}_{eff}^2=2{\theta}^2$. This statement can be made precise using the coherent states approach \cite{stern}. 

On all noncommutative spaces  it is always possible to map operators $\hat{O}$ to fields $O(x)$ using the so-called Weyl map. The pointwise multiplication of operators will be replaced by a star product while traces will be replaced by ordinary integrals . For example on the noncommutative Moyal-Weyl plane the $U(1)$ action (\ref{action1}) can be rewritten in this language as follows
\begin{eqnarray}
S_{\theta}=\frac{1}{4g^2}\int d^2x {F}_{ij}^2~,~F_{ij}={\partial}_iA_j-{\partial}_jA_i+i\{A_i,A_j\}_{*}.\label{action3}
\end{eqnarray} 
\section{The Noncommutative $U(1)$ Theory In $4-$Dimensions}

The action in higher dimensions is similar to (\ref{action3}). In ${\bf R}^d_{\theta}={\bf R}^{d-2}{\times}{\bf R}^2_{\theta}$ we have the commutation relations 
\begin{eqnarray}
[x_{\mu},x_{\nu}]=0~,~[x_{\mu},x_{i}]=0~,~[\hat{x}_i,\hat{x}_j]=i{\theta}^2{\epsilon}_{ij}~,\mu,\nu=1,...,d-2~,~i,j=d-1,d.\label{cond22}
\end{eqnarray}
For simplicity we are only considering minimal noncommutativity where only two spatial coordinates fail to commute. In order to maintain unitarity of the quantum  theory we are also assuming that the time direction lies in the commutative submanifold . The covariant derivatives in this case are given by $\hat{D}_{\mu}=-i{\partial}_{\mu}+\hat{A}_{\mu}$ , $\hat{D}_{i}=-\frac{1}{{\theta}^2}({\epsilon}^{-1})_{ij}\hat{x}_j+\hat{A}_i$ and the $U(1)$ action reads exactly like (\ref{action3}) where the star product is now given in terms of the commutation relations (\ref{cond22}).

This action can be reexpressed back in terms of operators as follows

\begin{eqnarray}
S_{\theta}
&{\equiv}&\frac{{\theta}^2}{4g^2}\int d^{d-2}xTr\hat{F}_{\mu
\nu}^2+\frac{{\theta}^2}{2g^2}\int
d^{d-2}x\sum_{i=d-1}^dTr\hat{F}_{\mu
i}^2+\frac{{\theta}^2}{4g^2}\int
d^{d-2}x\sum_{i,j=d-1}^dTr\hat{F}_{ij}^2.\label{action5}
\end{eqnarray}
In above we have deliberately used the fact that we can replace
the integral over the noncommutative directions $x_{d-1}$ and $x_d$
by a trace over an infinite dimensional Hilbert space by using the
Weyl Map. By doing this we have
therefore also replaced the underlying star product of functions
by pointwise multiplication of operators. The trace $Tr$ 
is thus associated with the two noncommutative
coordinates $x_{d-1}$ and $x_d$ . The model looks very much like 
a $U(\infty)$ gauge theory
on ${\bf R}^{d-2}$ with a Higgs particle in the adjoint of the
group.

In the remainder of this section we will confine ourselves to $4-$dimensions. From equation (\ref{action5}) we can see that for each point of the $2-$dimensional commutative
${\bf R}^{2}$ the above action  is
an infinite dimensional matrix model.  It can
be regularized if
we approximate the noncommutative plane by a fuzzy
sphere in exactly  the same way as before. The regularized action reads \cite{mine}
\begin{eqnarray}
S_{\theta;L}=\frac{1}{4{\lambda}^2}\int d^{2}x Tr_L{\cal
F}_{\mu \nu}^2-\frac{1}{2{\lambda}^2}\int
d^{2}x\sum_{a=1}^3Tr_L[{\cal
D}_{\mu},D_a]^2-\frac{1}{4{\lambda}^2}\int d^{2}xV(D_a).\label{action6}
\end{eqnarray}
$D_a$ are $(L+1){\times}(L+1)$ matrices which are fields on ${\bf R}^2$ and satisfy
\begin{eqnarray}
D_a^2=\frac{|L|^2}{R^2}.\label{cons}
\end{eqnarray}
The potential term is
\begin{eqnarray}
V(D_a)=Tr_L[D_a,D_b]^2- \frac{2i}{R}{\epsilon}_{abc}
Tr_L[D_a,D_b]D_c- \frac{2}{R^4}(L+1)|L|^2,\label{poten}
\end{eqnarray}
while the coupling constant is $
{\lambda}^2=\frac{g^2(L+1)}{R^2}$ where $g$ is the coupling constant on the noncommutative space ${\bf R}^4_{\theta}$ . 
\begin{eqnarray}
{\cal F}_{\mu \nu}=i[{\cal D}_{\mu},{\cal D}_{\nu}]~,~{\cal D}_{\mu}=-i{\partial}_{\mu}+{\cal A}_{\mu}. \label{def}
\end{eqnarray}
${\cal A}_{\mu}$ are $(L+1){\times}(L+1)$ matrices which are fields on ${\bf R}^2$ . This is clearly a
$U(L+1)$ gauge theory with adjoint matter , i.e the original
noncommutative degrees of freedom are traded for ordinary color
degrees of freedom. The
field ${\cal A}_{\mu}$ can be separated into a $U(1)$ gauge
field  $a_{\mu}$ and an $SU(L+1)$ gauge field $A_{\mu}$ as follows
\begin{eqnarray}
{\cal A}_\mu(x)=a_{\mu}(x){\bf 1}+A_{\mu}(x)~,~A_{\mu}(x)=A_{\mu
A}(x)T_A.
\end{eqnarray}
Similarly we write
\begin{eqnarray}
D_a={n}_{a}+{\Phi}_a~,~{\Phi}_a={\Phi}_{aA}T_A,
\end{eqnarray}
Under gauge transformations $n_a$ is a singlet while 
${\Phi}_a$ transforms in the adjoint representation of the non-abelian
group $SU(L+1)$ . These are ``scalars'' with
respect to the commutative directions of ${\bf R}^4_{\theta}={\bf R}^2_{\theta}{\times}{\bf R}^2$. The abelian $U(1)$ field $a_{\mu}$ is found from  the action to be free and thus it can be integrated out . The non-abelian $SU(L+1)$ field is seen to be defined on a two dimensional
spacetime and thus it can also be integrated out if one uses the
light-cone gauge. To this end we rotate first to
Minkowski  signature then we fix the $SU(L+1)$ symmetry by going
to the light-cone gauge given by 
\begin{eqnarray}
A_{-}=0{\Leftrightarrow}A_1=A_2=\sqrt{2}{\lambda}A_{+}.
\end{eqnarray}
The integral over the $A_{+}$ field becomes Gaussian and thus it can be easily done 
. It gives a non-local Coulomb
interaction between the ${\Phi}_{aC}$ fields. We define 
\begin{eqnarray}
{\Delta}_{AB}(x,y)=-\frac{{\delta}_{AB}}{2}|x_{-}-y_{-}|{\delta}(x_{+}-y_{+})~,~f_{ABC}({\partial}_{-}{\Phi}_{aA}){\Phi}_{aC}{\equiv}(\vec{\Phi}_a{\times}_L{\partial}_{-}\vec{\Phi}_a)_B\nonumber
\end{eqnarray}
where ${\Delta}_{AB}$ is clearly the propagator of the ${\Phi}_{aC}$ fields and then write the final result  in the form
\begin{eqnarray}
\hat{S}_{\theta;L}&=&\frac{N}{2{\lambda}^2}\int
d^{2}x({\partial}_{\mu}n_a)({\partial}^{\mu}n_a)+\frac{1}{4{\lambda}^2}\int
d^{2}x({\partial}_{\mu}{\Phi}_{aA})({\partial}^{\mu}{\Phi}_{aA})\nonumber\\
&-&\frac{1}{4{\lambda}^2}\int d^2xd^2y
(\vec{\Phi}_a{\times}_L{\partial}_{-}\vec{\Phi}_{a})_A(x){\Delta}_{AB}(x,y)(\vec{\Phi}_b{\times}_L{\partial}_{-}\vec{\Phi}_{b})_B(y)-\frac{1}{4{\lambda}^2}\int
d^{2}xV({\Phi}_a).\nonumber\\
\end{eqnarray}
The constraint $D_aD_a=\frac{|L|^2}{R^2}$ 
can be rewritten in the form 
\begin{eqnarray}
&&{n}^2_{a}+\frac{1}{2(L+1)}{\Phi}_{aA}^2=\frac{|L|^2}{{R}^2}~,~{n}_{a}{\Phi}_{aC}+\frac{1}{4}d_{ABC}{\Phi}_{aA}{\Phi}_{aB}=0,
\end{eqnarray}
 From the
structure of this constraint and from the action we
can see that the field $n_a$ appears at most quadratically. The corresponding path integral can be done exactly in the large $L$ limit and one obtains 
( with ${\chi}_{aA}=\frac{R}{|L|}\frac{1}{\sqrt{2N}}{\Phi}_{aA}$ ) the reduced action \cite{mine} 
\begin{eqnarray}
\bar{S}_{\theta;L}&=&\frac{1}{4{\bar{\lambda}}^2}\int
d^{2}x({\partial}_{\mu}{\chi}_{aA})({\partial}^{\mu}{\chi}_{aA})-\frac{|L|^2(L+1)}{2{\bar{\lambda}}^2{R}^2}\int
d^2x \bar{V}({\chi}_a)~,~{\bar{\lambda}}^2=\frac{g^2}{2|L|^2}
\end{eqnarray}
where
\begin{eqnarray}
\bar{V}({\chi}_a)&=&\int d^2y
(\vec{\chi}_a{\times}_L{\partial}_{-}\vec{\chi}_{a})_A(x){\Delta}_{AB}(x,y)(\vec{\chi}_b{\times}_L{\partial}_{-}\vec{\chi}_{b})_B(y)+Tr_L[{\chi}_a,{\chi}_b]^2\nonumber\\
&-&\frac{2i}{|L|\sqrt{2(L+1)}}{\epsilon}_{abc}Tr_L[{\chi}_a,{\chi}_b]{\chi}_c\nonumber-\frac{1}{2|L|^2(L+1)}.
\end{eqnarray}
The fields ${\chi}_{aA}$ satisfy now the constraints 
\begin{eqnarray}
{\chi}_{aA}^2=1~,~d_{ABC}{\chi}_{aA}{\chi}_{aB}=-\frac{2e_aR}{\sqrt{2|L|^2(L+1)}}{\chi}_{aC},
\end{eqnarray}
where $e_a$ is an arbitrary constant vector in ${\bf R}^3$ \cite{mine}.
Since
$R^2={\theta}^2|L|^{2q}$ the overall coupling in front of the potential
$\bar{V}$ behaves as
\begin{eqnarray}
\frac{|L|^2(L+1)}{2\bar{\lambda}^2R^2}{\sim}\frac{1}{{\bar{\lambda}}^2{\theta}^2}(\frac{L}{2})^{3-2q}
\end{eqnarray}
Thus for all scalings with $q>\frac{3}{2}$ this potential
can be neglected compared to the kinetic term. The fuzzy theory for these 
scalings becomes a theory living on a noncommutative plane with
effective deformation parameter 
\begin{eqnarray}
{\theta}_{eff}^2{\sim}2{\theta}^2(\frac{L}{2})^{2q-1}
\end{eqnarray}
We are therefore probing the strong
noncommutativity region of the Moyal-Weyl model . The 
partition function in this case is given by
\begin{eqnarray}
Z=\int {\cal D}J{\cal D}J_C e^{i \int d^2x J }
exp\bigg({-\frac{3}{2}TRlogD }
\bigg)exp\bigg(-{\vec{e}^2{\theta}^2|L|^{2(q-\frac{3}{2})}\int
d^2x d^2y J_A(x)D^{-1}_{AB}(x,y)J_B(y)}\bigg)\nonumber
\end{eqnarray}
where $D(=D_{AB}(x,y))$ is the Laplacian
\begin{eqnarray}
D_{AB}(x,y)={\delta}^2(x-y)\bigg(-\frac{1}{4{\bar{\lambda}}^2}{\partial}^2{\delta}_{AB}+iJ{\delta}_{AB}+iJ_Cd_{ABC}\bigg).
\end{eqnarray}
At this stage it is obvious that in the large $L$ limit only
configurations where $J_A=0$ are relevant and thus one ends up
with the partition function

\begin{eqnarray}
Z=\int {\cal D}J exp\bigg({\frac{i}{4\bar{\lambda}^2} \int d^2x
J-\frac{M}{2}\int d^2x <x|log\big(-{\partial}^2+iJ\big)|x>}\bigg).
\end{eqnarray}
This is exactly the partition function of an $O(M)$ non-linear
sigma model in the limit $M{\longrightarrow}{\infty}$ with
$\bar{\lambda}^2M$ held fixed equal to $\bar{\lambda}^2M=6g^2$ where
$M=3(N^2-1)=3L(L+2)$ . All terms in the exponent are now of the same order $M$ and thus
the model can  be solved using steepest descents . For example we can derive the beta function \cite{sigma}
\begin{eqnarray}
{\beta}(g_r)=\mu\frac{{\partial}g_r}{{\partial}\mu}=-\frac{3}{\pi}g_r^3.\label{beta1}
\end{eqnarray}
This result agrees nicely with the  one-loop
calculation of the beta function of  $U(1)$ theory on the Moyal-Weyl Plane \cite{nekrasov}. The crucial
difference  is the fact that this result is
exact to all orders in $\bar{\lambda}^2M=6g^2$ and
thus it is intrinsically nonperturbative \cite{sigma}. 
\section{$U(n)$ Gauge Theory and The Presnajder-Steinacker Action}
$U(n)$ gauge theory on ${\bf R}^{d-2}{\times}{\bf R}^2_{\theta}$ is given by the action
\begin{eqnarray}
S_{\theta}=\frac{1}{4g^2}\int d^dx \sum_{A,B=1}^dtr_n{F}_{AB}^2=\frac{{\theta}^2}{4g^2}\int d^{d-2}xTrtr_n\hat{F}_{AB}^2~,~A,B=1,...,d,\label{action7}
\end{eqnarray} 
with $F_{AB}={\partial}_AA_B-{\partial}_BA_A+i\{A_A,A_B\}_{*}$ and $\hat{F}_{AB}=\hat{{\partial}}_A\hat{A}_B-\hat{{\partial}}_B\hat{A}_A+i[\hat{A}_A,\hat{A}_B]$  . The regularized theory can be obtained in the same way as before and one ends up with the equations (\ref{action6}), (\ref{cons}) , (\ref{poten}) and (\ref{def}) with the only replacement $Tr_L{\longrightarrow}Tr_N=Tr_Ltr_n$. This is clearly a $U(N+1){\equiv}U(n(L+1))$ gauge theory which has also the interpretation of being a $U(n)$ gauge theory on ${\bf R}^{d-2}{\times}{\bf S}^2_L$ . The corresponding action can be simplified further if one uses the following trick due to Presnajder \cite{private} and Steinacker \cite{steinacker}.

The $3-$matrix action
$S_{L,R}$ given in equation (\ref{action2}) togther with the constraint (\ref{constraint1}) can be derived from a much simpler $1-$matrix model . To
this end we  introduce Pauli matrices ${\sigma}_a$ and we write the operator
\begin{eqnarray}
\bar{\Phi}=(\frac{1}{2}+{\sigma}_aL_a){\otimes}{\bf 1}_n.
\end{eqnarray}
This is a $2N-$dimensional matrix. It is a trivial exrecise to check that
$\bar{\Phi}=(j(j+1)-(\frac{L+1}{2})^2){\otimes}{\bf 1}_n$ where $j$ is the eigenvalue of the operator $\vec{J}=\vec{L}+\frac{\vec{\sigma}}{2}$  which takes the two values $\frac{L+1}{2}$ and $\frac{L-1}{2}$ . The eigenvalues of $\bar{\Phi}$ are therefore $\frac{L+1}{2}$ with multiplicity $n(L+2)$ and $-\frac{L+1}{2}$ with multiplicity $nL$ . As it turns out this matrix $\bar{\Phi}$ can be obtained as a classical configuration of the following $2N-$dimensional $1-$matrix action
\begin{eqnarray}
S[\Phi]=\frac{1}{4g^2R^2}\frac{1}{L+1}Tr_{N}tr_2\bigg[\frac{(L+1)^4}{16}-\frac{(L+1)^2}{2}{\Phi}^2+{\Phi}^4\bigg].
\end{eqnarray}
Indeed the equations of motion derived from this action reads
\begin{eqnarray}
{\Phi}({\Phi}^2-\frac{(L+1)^2}{4})=0.\nonumber
\end{eqnarray}
 It is easy to see that $\bar{\Phi}$ solves this equation of motion and that the value of the action in this configuration  is identically zero , i.e $S[\Phi=\bar{\Phi}]=0$ . In general and in terms of the eigenvalues ${\phi}_i$ of $\Phi$ this equation reads ${\phi}_i({\phi}_i^2-\frac{(L+1)^2}{4})=0$ which means that ${\phi}_i=0,+\frac{L+1}{2},-\frac{L+1}{2}$ , i.e classical configurations are matrices of eigenvalues $0$ , $+\frac{L+1}{2}$ and $-\frac{L+1}{2}$ with corresponding multiplicities   $n_{0}$ , $n_{+}$ and $n_{-}$ respectively which must clearly add up to $n_0+n_{+}+n_{-}=2n(L+1)$. The action for each zero eigenvalue ${\phi}_i=0$ is given by $S[{\phi}_i=0]=\frac{1}{4g^2R^2}\frac{1}{L+1}\frac{(L+1)^4}{16}$ which is suppressed in the large $L$ limit and thus these stationary points do not contribute in the large $L$ limit.Expanding around the vacuum $\bar{\phi}$ by writing 
\begin{eqnarray}
\Phi=\frac{1}{2}+\rho +R{\sigma}_aD_a
\end{eqnarray}
 where $\rho$ and $D_a$ are $N{\times}N$ matrices will immediately lead to the action $S_{L,R}$ if one  also imposes the condition \cite{steinacker}
\begin{eqnarray}
\rho=0.\label{re}
\end{eqnarray}
Indeed we find explicitly 
\begin{eqnarray}
S[\Phi=\frac{1}{2} +R{\sigma}_aD_a]=\frac{1}{4g^2}\frac{R^2}{L+1}Tr_Ltr_n\bigg[F_{ab}^2+2(D_a^2-\frac{|L|^2}{R^2})^2\bigg]~,~F_{ab}=i[D_a,D_b]+\frac{1}{R}{\epsilon}_{abc}D_c.
\end{eqnarray}
The second term in this action can be shown to implement exactly the constraint (\ref{constraint1}) as we want. The $U(n)$ action on ${\bf R}^{d-2}{\times}{\bf S}^2_{L}$ can then be taken ( without any further constraint ) to be

\begin{eqnarray}
S_{\theta;L}=\frac{1}{4{\lambda}^2}\int d^{d-2}x Tr_N{\cal
F}_{\mu \nu}^2-\frac{1}{2{\lambda}^2}\int
d^{d-2}x\sum_{a=1}^3Tr_N[{\cal
D}_{\mu},D_a]^2-\frac{1}{4{\lambda}^2}\int d^{d-2}xW(D_a)
\end{eqnarray}
where the potential reads now as follows 
\begin{eqnarray}
W(D_a)=-Tr_N\big(F_{ab}^2+2(D_a^2-\frac{|L|^2}{R^2})^2\big).
\end{eqnarray}
In terms of the scalar field $\Phi$ we can rewrite this action as follows 
\begin{eqnarray}
S_{\theta;L}&=&\frac{1}{4{\lambda}^2}\int d^{d-2}x Tr_N{\cal
F}_{\mu \nu}^2+\frac{1}{4{\lambda}^2R^2}\int d^{d-2}xTr_N[{\cal D}_{\mu},{\Phi}_{ij}]^{+}[{\cal D}_{\mu},{\Phi}_{ij}]\nonumber\\
&+&\frac{1}{4R^4{\lambda}^2}\int d^{d-2}xTr_N\big({\Phi}_{ij}{\Phi}_{jk}{\Phi}_{kl}{\Phi}_{li}-\frac{(L+1)^2}{2}{\Phi}_{ij}{\Phi}_{ji}+\frac{(L+1)^4}{16}\big).\label{gg}
\end{eqnarray}
Recall that ${\Phi}$ is a $2{\times}2$ matrix where each component ${\Phi}_{ij}$ is an $N{\times}N$ matrix . Under $U(N)$ gauge transformations  each of these components transforms covariantly . In deriving the above result we have used the Fierz identity $({\sigma}_a)_{ij}({\sigma}_a)_{kl}=2({\delta}_{il}{\delta}_{kj}-\frac{1}{2}{\delta}_{ij}{\delta}_{kl})$ as well as the constraint (\ref{re}) which we can write in the equivalent form
\begin{eqnarray}
{\Phi}_{11}+{\Phi}_{22}=1.\label{ggg}
\end{eqnarray}
The action (\ref{gg}) is essentially a Georgi-Glashow model with several scalar fields $\frac{{\Phi}_{ij}}{R}$ in the adjoint representation of the group which are restricted to satisfy the constraint (\ref{ggg}). In other words $U(n)$ gauge theory on ${\bf R}^{d-2}{\times}{\bf R}^2_{{\theta}_{eff}}$ can be approximated using a fuzzy sphere of matrix size $L+1$ and radius $R$ by a sequence of Georgi-Glashow models given by (\ref{gg})+(\ref{ggg}) with increasing $L$ and $R$ . The gauge groups are seen to be given by $U(n(L+1))$ while the coupling constants are given by ${\lambda}^2{\sim}\frac{g^2}{{\theta}_{eff}^2}$ where $g^2$ is the coupling constant on ${\bf R}^{d-2}{\times}{\bf R}^2_{{\theta}_{eff}}$.  The noncommutativity parameter on ${\bf R}^{d-2}{\times}{\bf R}^2_{{\theta}_{eff}}$ is found to be given by ${\theta}^2_{eff}{\sim}2{\theta}^2(\frac{L}{2})^{2q-1}$ where ${\theta}^2=R^2/|L|^{2q}$ is always kept fixed. Clearly the quantum theory depends on the way we take the limit. We defer the study of these models to a future correspondence. 

\section{Conclusion} 

In this article we have considered gauge theory on ${\bf R}^{d-2}{\times}{\bf R}^2_{\theta}$ . We have regularized the two noncommuting directions by replacing them with a fuzzy sphere . This turns the noncommutative field theory into an ordinary commutative field theory amenable to the standard techniques of quantization and renormalization , etc. The non-trivial ingredient in this construction remains always  the definition of the limit which requires in our opinion further study . The $U(1)$ theory is seen in some scaling limit to correspond to an ordinary $2d$ non-linear sigma model thus allowing us to derive the beta function of the theory . The result agrees with perturbation theory but the question remains what happens in other scaling limits. Higher $U(n)$ in higher dimensions are found to be classically equivalent to  a sequence of Georgi-Glashow models defined on the commutative submanifold. Their quantum properties will however be studied elsewhere.

\bibliographystyle{unsrt}

\end{document}